\documentclass[preprint,showpacs,nofootinbib]{revtex4}

\begin{document}

\title{On the black hole limit of rotating fluid bodies in equilibrium\footnote{Fondly dedicated to Gernot Neugebauer on the occasion of his 65\raisebox{1ex}{th} birthday.}}

\author{Reinhard Meinel}
\affiliation{
University of Jena, Theoretisch-Physikalisches Institut,\\ Max-Wien-Platz 1, 
07743 Jena, Germany
}

\date{December 23, 2005}

\begin{abstract}
Recently, it was shown that the extreme Kerr black hole is the only candidate for a (Kerr) black hole limit of stationary and axisymmetric, uniformly rotating perfect fluid bodies with a zero temperature equation of state. In this paper, necessary and sufficient conditions for reaching the black hole limit are presented.
\end{abstract}

\pacs{04.20.Dw, 04.40.Dg, 04.70.Bw} 

\maketitle

\section{Introduction}
It was shown in \cite{meinel04} that a continuous sequence of stationary and axisymmetric, uniformly rotating perfect fluid bodies with a ``cold matter'' equation of state and finite baryonic mass can reach a (Kerr) black hole limit only if the relation
\begin{equation}
M=2\Omega J
\label{cond1}
\end{equation}
between the gravitational mass $M$, the angular velocity $\Omega$ and the angular momentum $J$ is satisfied in the limit.\footnote{We use units in which the speed of light as well as Newton's gravitational constant are equal to 1. Strictly speaking, $\Omega$ and $J$ are the components of angular velocity and angular momentum with respect to the axis of symmetry and can have either sign, corresponding to the two possible directions of rotation.} This result implies the impossibility of  black hole limits of non-rotating equilibrium configurations (cf.\ ``Buchdahl's inequality''), and moreover, since $\Omega$ must become equal to the ``angular velocity of the horizon'' 
\begin{equation}
\Omega^H= \frac{J}{2M^2\left[M+\sqrt{M^2-(J/M)^2}\,\right]}\, ,
\label{OmH}
\end{equation}
the relation
\begin{equation}
J=\pm M^2,
\end{equation}
characteristic of an {\it extreme} Kerr black hole, must hold in the limit. The possibility of such a limit was first demonstrated for  infinitesimally thin disks -- numerically by Bardeen and Wagoner \cite{BW71} and analytically by Neugebauer and Meinel \cite{NM95}, see also \cite{meinel02}. Further numerical examples, for genuine fluid bodies, were provided by the ``relativistic Dyson rings'' \cite{AKM4} and their generalizations \cite{FHA}.

The aim of the present paper is to show that condition (\ref{cond1}) is not only necessary, but also sufficient for reaching the black hole limit. An equivalent condition will turn out to be the statement that a zero angular momentum photon emitted from the fluid's surface suffers an infinite redshift.

\section{Basic relations} In the following, some basic relations for rotating fluids in equilibrium are provided, for more details see, for example, \cite{HS, BW71, Thorne, li, fi, NM03}. The four-velocity of the fluid must point in the direction of a
linear combination of the two commuting Killing vectors
$\xi=\partial/\partial t$ and $\eta=\partial/\partial \phi$
corresponding to stationarity and axisymmetry:
\begin{equation}
u^i=e^{-V}(\xi^i+\Omega\,\eta^i),\quad \Omega=\mbox{\rm constant.}
\label{four}
\end{equation}
The Killing vector $\xi$ is fixed by the normalization
$\xi^i\xi_i\to -1$ at spatial infinity (we assume asymptotic
flatness).\footnote{The spacetime signature is chosen to be ($-+++$).} The orbits of the spacelike Killing vector $\eta$ are
closed and $\eta$ is zero on the axis of symmetry. The constant of eq.\ (\ref{four}),
$\Omega=u^\phi/u^t$, is the angular velocity of the fluid body with
respect to infinity. Using $u^iu_i=-1$, the factor $e^{-V}=u^t$ is given by
\begin{equation}
(\xi^i+\Omega\,\eta^i)(\xi_i+\Omega\,\eta_i)=-e^{2V}.
\label{V}
\end{equation}
The energy-momentum tensor is 
\begin{equation}
T_{ik}=(\epsilon+p)\,u_iu_k+p\,g_{ik},
\end{equation}
where the mass-energy density $\epsilon$ and the pressure $p$ are
related by a ``cold'' equation of state, $\epsilon=\epsilon(p)$,
following from \noindent
\begin{equation}
p=p(\rho,T),\quad \epsilon=\epsilon(\rho,T)
\end{equation}
for $T=0$, where $\rho$ is the baryonic mass-density and $T$ the temperature.
The specific enthalpy
\begin{equation}
h=\frac{\epsilon+p}{\rho}
\label{h}
\end{equation}
can be calculated from $\epsilon(p)$ via the thermodynamic relation
\begin{equation}
dh=\frac{1}{\rho}\,dp \qquad (T=0)
\end{equation}
leading to
\begin{equation} 
\frac{dh}{h}=\frac{dp}{\epsilon+p} \Rightarrow
h(p)=h(0)\,\exp\left[\int\limits_0^p\frac{dp'}{\epsilon(p')+p'}\right].
\label{enth}
\end{equation}
Note that $h(0)=1$ in most cases. For our
purposes, however, it is sufficient to assume
$0<h(0)<\infty$. From ${T^{ik}}_{;k}=0$ we obtain
\begin{equation}
h(p)\,e^V=h(0)\,e^{V_0}={\rm constant},
\label{cc}
\end{equation}
where $V_0$, the constant surface value (corresponding to $p=0$)
of the function $V$ defined in (\ref{V}), is related to the
relative redshift $z$ of zero angular momentum
photons\footnote{Zero angular momentum means $\eta_ip^i=0$ ($p^i$: four-momentum of
the photon).} emitted from the surface of the fluid and received
at infinity:
\begin{equation}
z=e^{-V_0}-1.
\end{equation}
Equilibrium models, for a given equation of state, are fixed by
two parameters, for example $\Omega$ and $V_0$. (When we discuss a ``sequence'' of solutions, what is meant is a curve in the two-dimensional parameter space.) The gravitational mass and the angular momentum can be calculated by
\begin{equation}
M=2\int\limits_{\Sigma}(T_{ik} - \frac{1}{2}Tg_{ik})n^i\xi^kd
{\cal V}, \quad J=-\int\limits_{\Sigma}T_{ik}\,n^i\eta^kd{\cal V},
\label{MJ}
\end{equation}
where $\Sigma$ is a spacelike hypersurface ($t={\rm constant}$)
with the volume element $d{\cal V}=\sqrt{^{(3)}g}\,d^3x$ and the
future pointing unit normal $n^i$, see for example \cite{Wald}. The baryonic
mass $M_0$ corresponding to the local conservation law $(\rho
u^i)_{;i}=0$ is
\begin{equation}
M_0=-\int\limits_{\Sigma}\rho\,u_i\,n^id{\cal V}.
\label{M0}
\end{equation}
Note that nearby equilibrium configurations with the same equation
of state are related by \cite{HS, BW71}
\begin{equation}
\delta M=\Omega\, \delta J + \mu\,\delta M_0, \qquad \mu=h(0)\,e^{V_0}.
\label{var}
\end{equation}
The parameter $\mu$ (``chemical potential'') represents the specific injection energy of zero angular momentum baryons.

\section{Conditions for a black hole limit}
A combination of (\ref{MJ}) and (\ref{M0}) leads to the formula
\begin{equation}
M=2\Omega J+\int\frac{\epsilon+3p}{\rho}\,e^VdM_0,
\end{equation}
cf.\ equation (II.28) in \cite{BW71}. With (\ref{h}) and (\ref{cc})
we get 
\begin{equation}
M=2\Omega J+
h(0)\,e^{V_0}\int\frac{\epsilon+3p}{\epsilon+p}\,dM_0.
\label{mass}
\end{equation}
Since $1\le (\epsilon+3p)/(\epsilon+p)\le 3$ (we assume $\epsilon$ and $p$ to be non-negative), condition (\ref{cond1}) is equivalent to\footnote{We assume $0< M_0 < \infty$.}
\begin{equation}
V_0\to -\infty \quad (z\to\infty).
\end{equation} 
We now want to show that this condition is not only necessary (as discussed in \cite{meinel04}), but also sufficient for approaching a black hole limit. 

Because of (\ref{V}) and (\ref{cc}), the surface of the fluid is
characterized in general by
\begin{equation}
\chi^i\chi_i=-e^{2V_0},
\quad \chi^i\equiv \xi^i+\Omega\eta^i.
\end{equation}
The Killing vector $\chi^i$ is tangential to the hypersurface $\cal{H}$ generated by the timelike world lines of the fluid elements of the surface of the body with four-velocity $u^i=e^{-V_0}\chi^i$, see (\ref{four}). Each of the Killing vectors $\xi^i$ and $\eta^i$ must itself be tangential to  $\cal{H}$ because of the symmetries of the spacetime.
In the limit $V_0\to -\infty$, we approach a situation in which $\chi^i$ becomes null on $\cal{H}$:
\begin{equation}
\chi^i\chi_i\to 0.
\label{null}
\end{equation}
Moreover, with the reasonable assumption\footnote{The condition $-\xi^i u_i\le 1$ ensures that a particle resting on the surface of the fluid is (at least marginally) bound, i.e.\ cannot escape to infinity on a geodesic;
$-\xi^i u_i\ge 0$ follows from $-\xi^i u_i=-(\chi^i-\Omega\eta^i)u_i=e^{V_0}+\Omega\eta^i u_i$, since $\eta^i u_i$ will always have the same sign as $\Omega$ ($\eta^i u_i = 0$ on the axis, of course).}
\begin{equation}
0\le -\xi^i u_i \le 1,
\end{equation}
we find that $\chi^i$ also becomes orthogonal to $\xi^i$ (and thus to $\eta^i$) on  $\cal{H}$ in the limit:
\begin{equation}
\chi^i\xi_i\to 0, \quad \chi^i\eta_i\to 0.
\label{ortho}
\end{equation}
Together with the orthogonal transitivity\footnote{The conditions of the theorem by Kundt and Tr\"umper \cite{Ku} are satisfied.} of the spacetime, $\chi^i$ therefore becomes orthogonal to three linearly independent tangent vectors at each point of $\cal{H}$, i.e.\ normal to $\cal{H}$. Because of (\ref{null}), we thus approach a situation in which $\cal{H}$ is a null hypersurface and satisfies all defining conditions for a horizon of a stationary (and axisymmetric) black hole with
$\Omega$ being the angular velocity of the horizon, see \cite{Carter73}. According to the black hole uniqueness theorems (see \cite{HE, heusler} and also \cite{NM03}) we conclude that (outside the horizon) the Kerr metric with $|J| \le M^2$ results\footnote{Note that the black hole uniqueness proof by construction
given in \cite{NM03} can be extended to the case in which the horizon is degenerate,
leading to $|J| = M^2$. This will be shown in a future publication.}.  
Then, with (\ref{cond1}), we are necessarily led to the case $|J| = M^2$.  Therefore, the metric of an extreme Kerr black hole (outside the horizon) results, whenever a sequence of fluid bodies admits a limit 
$V_0\to -\infty$. 

\section{Discussion}
The special properties of the {\it extreme} Kerr metric with its degenerate horizon and the infinitely long ``throat region'' allow for the existence of a black hole limit independent of the fluid body's topology. Indeed, such a limit was found numerically for bodies of toroidal topology \cite{AKM4}. Strictly speaking, there is not yet a horizon in the limit. Instead, a separation of an ``inner'' and an ``outer'' world occurs. The ``inner world'' contains the fluid body and is not asymptotically flat, but approaches the ``extreme Kerr throat geometry'' \cite{BH99} at spatial infinity. The ``outer world'' is given by the $r>M$ part of the extreme Kerr metric, where $r$ is the radial Boyer-Lindquist coordinate. Note that the horizon as well as the ``throat region'' are characterized by $r=M$. Here, the whole ``inner world'' corresponds to $r=M$. It should be mentioned in this connection that the conditions (\ref{null}) and (\ref{ortho}) are also satisfied {\it inside} the fluid as $V_0\to -\infty$, cf.\ (\ref{enth}) and (\ref{cc}). More details can be found in \cite{BW71, meinel02, AKM4}. 

It is interesting to note that a similar separation of spacetimes has been observed for some limiting solutions of the static, spherically symmetric Einstein-Yang-Mills-Higgs equations \cite{BMK, BFM} leading to the extreme Reissner-Nordstr\"om metric in the ``outer world''. 

As discussed in \cite{Bardeen73}, the slightest dynamical perturbation will lead to a genuine black hole. Therefore, it is tempting to continue a sequence of fluid bodies beyond the black hole limit as a sequence of Kerr black holes and to discuss the transition from the ``normal matter state'' to the ``black hole state'' as a (one-way) phase transition \cite{NM93}. From the exterior point of view, this transition is continuous, i.e.\ all gravitational multipole moments change continuously. It is interesting to compare the mass formula (\ref{mass}) as well as the differential relation (\ref{var}) with the black hole formulas \cite{Smarr, BCH}
\begin{equation}
M=2\Omega^H J + \frac{\kappa}{4\pi} A,
\end{equation}
\begin{equation}
\delta M= \Omega^H \delta J + \frac{\kappa}{8\pi} \delta A,
\end{equation} 
where $\kappa$ denotes the ``surface gravity'' and $A$ the area of the horizon. In the two-dimensional parameter space, the transition line is characterized by $M=2\Omega J=2\Omega^H J$. On the ``fluid side'' of this line, the parameter $\mu=h(0)e^{V_0}$ of the chemical potential vanishes ($V_0\to -\infty$), whereas $\kappa$ (related to the temperature in black hole thermodynamics) vanishes on the ``black hole side'' of the transition line ($\kappa = 0$ for extreme Kerr black holes). The quantities $M_0$ (baryonic mass of the fluid body) and $A$ (related to the black hole entropy) are defined in the corresponding regions of the parameter space only. This is consistent with $\mu\equiv 0$ in the black hole region and $T\equiv 0$ in the fluid region, i.e.\ $\mu$ and $T$ are continuous across the transition line. (An alternative interpretation was given in \cite{Neugebauer}.)

The parametric transitions from fluid bodies to black holes discussed here may be used as a starting-point for dynamical collapse investigations far from the spherically symmetric case. 
\begin{acknowledgments}
I would like to thank G.~Neugebauer, D.~Petroff, A.~Kleinw\"achter and H.~Labranche for valuable discussions. This work was supported in part by the Deutsche Forschungsgemeinschaft (DFG project SFB/TR7-B1).
\end{acknowledgments}


\begin{thebibliography}{23}
\expandafter\ifx\csname natexlab\endcsname\relax\def\natexlab#1{#1}\fi
\expandafter\ifx\csname bibnamefont\endcsname\relax
  \def\bibnamefont#1{#1}\fi
\expandafter\ifx\csname bibfnamefont\endcsname\relax
  \def\bibfnamefont#1{#1}\fi
\expandafter\ifx\csname citenamefont\endcsname\relax
  \def\citenamefont#1{#1}\fi
\expandafter\ifx\csname url\endcsname\relax
  \def\url#1{\texttt{#1}}\fi
\expandafter\ifx\csname urlprefix\endcsname\relax\def\urlprefix{URL }\fi
\providecommand{\bibinfo}[2]{#2}
\providecommand{\eprint}[2][]{\url{#2}}

\bibitem[{\citenamefont{Meinel}(2004)}]{meinel04}
\bibinfo{author}{\bibfnamefont{R.}~\bibnamefont{Meinel}},
  \bibinfo{journal}{Ann.\ Phys.\ (Leipzig)} \textbf{\bibinfo{volume}{13}},
  \bibinfo{pages}{600} (\bibinfo{year}{2004}).

\bibitem[{\citenamefont{Bardeen and Wagoner}(1971)}]{BW71}
\bibinfo{author}{\bibfnamefont{J.~M.} \bibnamefont{Bardeen}} \bibnamefont{and}
  \bibinfo{author}{\bibfnamefont{R.~V.} \bibnamefont{Wagoner}},
  \bibinfo{journal}{Astrophys.\ J.} \textbf{\bibinfo{volume}{167}},
  \bibinfo{pages}{359} (\bibinfo{year}{1971}).

\bibitem[{\citenamefont{Neugebauer and Meinel}(1995)}]{NM95}
\bibinfo{author}{\bibfnamefont{G.}~\bibnamefont{Neugebauer}} \bibnamefont{and}
  \bibinfo{author}{\bibfnamefont{R.}~\bibnamefont{Meinel}},
  \bibinfo{journal}{Phys.\ Rev.\ Lett.} \textbf{\bibinfo{volume}{75}},
  \bibinfo{pages}{3046} (\bibinfo{year}{1995}).

\bibitem[{\citenamefont{Meinel}(2002)}]{meinel02}
\bibinfo{author}{\bibfnamefont{R.}~\bibnamefont{Meinel}},
  \bibinfo{journal}{Ann.\ Phys.\ (Leipzig)} \textbf{\bibinfo{volume}{11}},
  \bibinfo{pages}{509} (\bibinfo{year}{2002}).

\bibitem[{\citenamefont{Ansorg et~al.}(2003)\citenamefont{Ansorg,
  Kleinw{\"a}chter, and Meinel}}]{AKM4}
\bibinfo{author}{\bibfnamefont{M.}~\bibnamefont{Ansorg}},
  \bibinfo{author}{\bibfnamefont{A.}~\bibnamefont{Kleinw{\"a}chter}},
  \bibnamefont{and} \bibinfo{author}{\bibfnamefont{R.}~\bibnamefont{Meinel}},
  \bibinfo{journal}{Astrophys.\ J.\ Lett.} \textbf{\bibinfo{volume}{582}},
  \bibinfo{pages}{L87} (\bibinfo{year}{2003}).

\bibitem[{\citenamefont{Fischer et~al.}()\citenamefont{Fischer, Horatschek, and
  Ansorg}}]{FHA}
\bibinfo{author}{\bibfnamefont{T.}~\bibnamefont{Fischer}},
  \bibinfo{author}{\bibfnamefont{S.}~\bibnamefont{Horatschek}},
  \bibnamefont{and} \bibinfo{author}{\bibfnamefont{M.}~\bibnamefont{Ansorg}},
  \bibinfo{journal}{Mon.\ Not.\ R.\ Astron.\ Soc.} \textbf{\bibinfo{volume}{364}},
  \bibinfo{pages}{943} (\bibinfo{year}{2005}).

\bibitem[{\citenamefont{Hartle and Sharp}(1967)}]{HS}
\bibinfo{author}{\bibfnamefont{J.~B.} \bibnamefont{Hartle}} \bibnamefont{and}
  \bibinfo{author}{\bibfnamefont{D.~H.} \bibnamefont{Sharp}},
  \bibinfo{journal}{Astrophys.\ J.} \textbf{\bibinfo{volume}{147}},
  \bibinfo{pages}{317} (\bibinfo{year}{1967}).

\bibitem[{\citenamefont{Thorne}()}]{Thorne}
\bibinfo{author}{\bibfnamefont{K.~S.} \bibnamefont{Thorne}}, \bibinfo{note}{in
  {\it Relativistic Astrophysics, Volume 1,} by Ya.\ B.\ Zel'dovich and I.\ D.\
  Novikov (The University of Chicago Press, Chicago, 1971), p.~264}.

\bibitem[{\citenamefont{Lindblom}(1992)}]{li}
\bibinfo{author}{\bibfnamefont{L.}~\bibnamefont{Lindblom}},
  \bibinfo{journal}{Phil.\ Trans.\ R.\ Soc.\ Lond.\ A}
  \textbf{\bibinfo{volume}{340}}, \bibinfo{pages}{353} (\bibinfo{year}{1992}).

\bibitem[{\citenamefont{Friedman and Ipser}(1992)}]{fi}
\bibinfo{author}{\bibfnamefont{J.~L.} \bibnamefont{Friedman}} \bibnamefont{and}
  \bibinfo{author}{\bibfnamefont{J.~R.} \bibnamefont{Ipser}},
  \bibinfo{journal}{Phil.\ Trans.\ R.\ Soc.\ Lond.\ A}
  \textbf{\bibinfo{volume}{340}}, \bibinfo{pages}{391} (\bibinfo{year}{1992}).

\bibitem[{\citenamefont{Neugebauer and Meinel}(2003)}]{NM03}
\bibinfo{author}{\bibfnamefont{G.}~\bibnamefont{Neugebauer}} \bibnamefont{and}
  \bibinfo{author}{\bibfnamefont{R.}~\bibnamefont{Meinel}},
  \bibinfo{journal}{J.\ Math.\ Phys.} \textbf{\bibinfo{volume}{44}},
  \bibinfo{pages}{3407} (\bibinfo{year}{2003}).

\bibitem[{\citenamefont{Wald}(1984)}]{Wald}
\bibinfo{author}{\bibfnamefont{R.~M.} \bibnamefont{Wald}},
  \emph{\bibinfo{title}{General Relativity}} (\bibinfo{publisher}{The
  University of Chicago Press}, \bibinfo{address}{Chicago},
  \bibinfo{year}{1984}).

\bibitem[{\citenamefont{Kundt and Tr{\"u}mper}(1966)}]{Ku}
\bibinfo{author}{\bibfnamefont{W.}~\bibnamefont{Kundt}} \bibnamefont{and}
  \bibinfo{author}{\bibfnamefont{M.}~\bibnamefont{Tr{\"u}mper}},
  \bibinfo{journal}{Z.\ Phys.} \textbf{\bibinfo{volume}{192}},
  \bibinfo{pages}{419} (\bibinfo{year}{1966}).

\bibitem[{\citenamefont{Carter}(1973)}]{Carter73}
\bibinfo{author}{\bibfnamefont{B.}~\bibnamefont{Carter}}, in
  \emph{\bibinfo{booktitle}{Black Holes, Les astres occlus}}, edited by
  \bibinfo{editor}{\bibfnamefont{C.}~\bibnamefont{DeWitt}} \bibnamefont{and}
  \bibinfo{editor}{\bibfnamefont{B.~S.} \bibnamefont{DeWitt}}
  (\bibinfo{publisher}{Gordon and Breach Science Publishers},
  \bibinfo{address}{New York}, \bibinfo{year}{1973}), p.~\bibinfo{pages}{57}.

\bibitem[{\citenamefont{Hawking and Ellis}(1973)}]{HE}
\bibinfo{author}{\bibfnamefont{S.~W.} \bibnamefont{Hawking}} \bibnamefont{and}
  \bibinfo{author}{\bibfnamefont{G.~F.~R.} \bibnamefont{Ellis}},
  \emph{\bibinfo{title}{The Large Scale Structure of Space-Time}}
  (\bibinfo{publisher}{Cambridge University Press},
  \bibinfo{address}{Cambridge, UK}, \bibinfo{year}{1973}).

\bibitem[{\citenamefont{Heusler}(1996)}]{heusler}
\bibinfo{author}{\bibfnamefont{M.}~\bibnamefont{Heusler}},
  \emph{\bibinfo{title}{Black Hole Uniqueness Theorems}}
  (\bibinfo{publisher}{Cambridge University Press},
  \bibinfo{address}{Cambridge, UK}, \bibinfo{year}{1996}).

\bibitem[{\citenamefont{Bardeen and Horowitz}(1999)}]{BH99}
\bibinfo{author}{\bibfnamefont{J.} \bibnamefont{Bardeen}} \bibnamefont{and}
  \bibinfo{author}{\bibfnamefont{G.~T.} \bibnamefont{Horowitz}},
  \bibinfo{journal}{Phys.\ Rev.\ D} \textbf{\bibinfo{volume}{60}},
  \bibinfo{pages}{104030} (\bibinfo{year}{1999}).
  
\bibitem[{\citenamefont{Bartnik and McKinnon}(1988)}]{BMK}
\bibinfo{author}{\bibfnamefont{R.}~\bibnamefont{Bartnik}}
  \bibnamefont{and} \bibinfo{author}{\bibfnamefont{J.}~\bibnamefont{McKinnon}},
  \bibinfo{journal}{Phys.\ Rev.\ Lett.} \textbf{\bibinfo{volume}{61}},
  \bibinfo{pages}{141} (\bibinfo{year}{1988}).
  
\bibitem[{\citenamefont{Breitenlohner et~al.}(1995)\citenamefont{Breitenlohner,
  Forg{\'a}cs, and Maison}}]{BFM}
\bibinfo{author}{\bibfnamefont{P.}~\bibnamefont{Breitenlohner}},
  \bibinfo{author}{\bibfnamefont{P.}~\bibnamefont{Forg{\'a}cs}},
  \bibnamefont{and} \bibinfo{author}{\bibfnamefont{D.}~\bibnamefont{Maison}},
  \bibinfo{journal}{Nucl.\ Phys.\ B} \textbf{\bibinfo{volume}{442}},
  \bibinfo{pages}{126} (\bibinfo{year}{1995}).

\bibitem[{\citenamefont{Bardeen}(1973)}]{Bardeen73}
\bibinfo{author}{\bibfnamefont{J.~M.} \bibnamefont{Bardeen}}, in
  \emph{\bibinfo{booktitle}{Black Holes, Les astres occlus}}, edited by
  \bibinfo{editor}{\bibfnamefont{C.}~\bibnamefont{DeWitt}} \bibnamefont{and}
  \bibinfo{editor}{\bibfnamefont{B.~S.} \bibnamefont{DeWitt}}
  (\bibinfo{publisher}{Gordon and Breach Science Publishers},
  \bibinfo{address}{New York}, \bibinfo{year}{1973}), p. \bibinfo{pages}{241}.

\bibitem[{\citenamefont{Neugebauer and Meinel}(1993)}]{NM93}
\bibinfo{author}{\bibfnamefont{G.}~\bibnamefont{Neugebauer}} \bibnamefont{and}
  \bibinfo{author}{\bibfnamefont{R.}~\bibnamefont{Meinel}},
  \bibinfo{journal}{Astrophys.\ J.\ Lett.} \textbf{\bibinfo{volume}{414}},
  \bibinfo{pages}{L97} (\bibinfo{year}{1993}).

\bibitem[{\citenamefont{Smarr}(1973)}]{Smarr}
\bibinfo{author}{\bibfnamefont{L.}~\bibnamefont{Smarr}},
  \bibinfo{journal}{Phys.\ Rev.\ Lett.} \textbf{\bibinfo{volume}{30}},
  \bibinfo{pages}{71} (\bibinfo{year}{1973}).

\bibitem[{\citenamefont{Bardeen et~al.}(1973)\citenamefont{Bardeen, Carter, and
  Hawking}}]{BCH}
\bibinfo{author}{\bibfnamefont{J.~M.} \bibnamefont{Bardeen}},
  \bibinfo{author}{\bibfnamefont{B.}~\bibnamefont{Carter}}, \bibnamefont{and}
  \bibinfo{author}{\bibfnamefont{S.~W.} \bibnamefont{Hawking}},
  \bibinfo{journal}{Commun.\ Math.\ Phys.} \textbf{\bibinfo{volume}{31}},
  \bibinfo{pages}{161} (\bibinfo{year}{1973}).

\bibitem[{\citenamefont{Neugebauer}(1998)}]{Neugebauer}
\bibinfo{author}{\bibfnamefont{G.}~\bibnamefont{Neugebauer}}, in
  \emph{\bibinfo{booktitle}{Black Holes: Theory and Observation}}, edited by
  \bibinfo{editor}{\bibfnamefont{F.~W.} \bibnamefont{Hehl}} \bibnamefont{and}
  \bibinfo{editor}{\bibfnamefont{C.}~\bibnamefont{Kiefer}}
  (\bibinfo{publisher}{Springer}, \bibinfo{address}{Berlin},
  \bibinfo{year}{1998}), p. \bibinfo{pages}{319}.

\end{thebibliography}
\end{document}